\documentclass[twocolumn,showpacs,showkeys,amsmath,amssymb,floatfix]{revtex4}
\usepackage{amsfonts}
\usepackage{amsmath}
\usepackage{graphicx}
\usepackage{subfigure}
\usepackage{dcolumn}
\usepackage{bm}
\usepackage{float} 
\usepackage{booktabs}
\usepackage[utf8]{inputenc}
\usepackage{stfloats}
\usepackage{graphicx,graphics,dcolumn,booktabs,bm}
\usepackage{makecell}
\usepackage{longtable,lscape}
\usepackage{txfonts}
\usepackage{overpic}
\usepackage{amssymb}
\usepackage{indentfirst}
\usepackage{epsfig}
\usepackage{feynmf}  
\usepackage{hyperref}
\usepackage{epstopdf}   
\usepackage{slashed}  
\usepackage{multirow}

\usepackage{ulem}
\usepackage[usenames,dvipsnames]{color}

\makeatletter
\newcommand{\figcaption}{\def\@captype{figure}\caption}
\newcommand{\tabcaption}{\def\@captype{table}\caption}

\newcommand{\Rmnum}[1]{\expandafter\@slowromancap\romannumeral #1@}
\def\hlinewd#1{%
  \noalign{\ifnum0=`}\fi\hrule \@height #1 \futurelet
   \reserved@a\@xhline}
\makeatother

\def\f(s){[(\alpha+\beta)m^2-\alpha\beta s]}

\newcommand{\jpsi}{J\kern-0.1em/\kern-0.1em\psi\kern0.03em}

\begin{document}

\title{Multi-hadron molecules: status and prospect}

\author{Tian-Wei Wu}
\affiliation{School of Fundamental Physics and Mathematical Sciences, Hangzhou Institute for Advanced Study, UCAS, Hangzhou 310024, China}
\affiliation{University of Chinese Academy of Sciences, Beijing 100049, China}

\author{Ya-Wen Pan}
\affiliation{School of
Physics,  Beihang University, Beijing 102206, China}

\author{Ming-Zhu Liu}
\affiliation{School of Space and Environment,  Beihang University, Beijing 102206, China}
\affiliation{School of
Physics,  Beihang University, Beijing 102206, China}

\author{Li-Sheng Geng}
\email[Corresponding author: ]{lisheng.geng@buaa.edu.cn}
\affiliation{School of
Physics,  Beihang University, Beijing 102206, China}
\affiliation{Beijing Key Laboratory of Advanced Nuclear Materials and Physics, Beihang University, Beijing 100191, China }
\affiliation{School of Physics and Microelectronics, Zhengzhou University, Zhengzhou, Henan 450001, China }

\begin{abstract}
Starting from 2003, the discovery of a large amount of the so-called exotic hadronic states, i.e., the $XYZ$ states, the pentaquark states as well as the tetraquark states, have not only revived studies of hadron spectroscopy, but also hinted at the existence of new multi-hadron states made of hadrons other than nucleons and hyperons. We briefly comment on some of the latest studies on multi-hadron molecules in the light and heavy flavor sectors and highlight what should be done in the future.     
\end{abstract}
\maketitle

To understand the fundamental particles of which our universe is composed and the underlying forces governing their motion has long been pursued by generations  of philosophers and scientists. Today, it is widely accepted that quarks and leptions are the most fundamental constituents of our visible universe, while the interactions among them are mediated by various force carriers, the photon for the electromagnetic interaction, the $W$ and $Z$ bosons for the weak interaction, and the gluons for the strong interaction. 

Quarks are colored objects and could not exist in isolation. According to the naive quark model proposed in 1964 by Gell-Mann and Zweig, three quarks form baryons and a pair of quark and antiquark constitute  mesons, which are collectively referred to as hadrons. The naive quark model has been very successful in explaining most if not all of the experimentally discovered baryons and mesons.
Nevertheless, the underlying theory of the strong interaction, ChromoDynamics (QCD), allows for the existence of hadronic states of more complicated structure --- the so-called exotic states. 
In general, exotic states can be categorized  into compact multiquark states, such as tetraquark states ($qq\bar{q}\bar{q}$) and pentaquark states ($qqqq\bar{q}$) and weakly-bound hadronic molecules that are composed of two or more conventional hadrons.  Considering the gluon degree of freedom, one expects the existence of glueballs consisting of only gluons and hybrid states consisting of both quarks and gluons.

Staring from 2003, a large amount of exotic hadronic states beyond the naive quark model have been observed, which ushered in a new era in hadron physics and studies of the non-perturbative strong interaction. Understanding the nature of these states has been a central issue both theoretically and experimentally since then~\cite{Brambilla:2019esw,Chen:2022asf}. For example, $X(3872)$ observed by the Belle Collaboration is still under debate regarding whether it is a conventional charmonium or a $D\bar{D}^*$ molecule. A similar state is the $D_{s0}^*(2317)$, which is located 160 MeV below the $P$-wave $c\bar{s}$ state  predicted in the naive quark model. In 2015 and 2019, the LHCb Collaboration observed four pentaquark states, $P_c(4380)$, $P_c(4312)$, $P_c(4440)$, and $P_c(4457)$. Given their masses and minimum quark contents $c\bar{c}qqq$, they qualify  as $\Sigma_{c}\bar{D}^
{(*)}$ molecules. In 2021, the LHCb Collaboration observed the first doubly charmed tetraquark state $T_{cc}^+(3875)$, which can be naturally explained as a $DD^*$  bound state.

Although a large amount of theoretical and experimental studies have been performed to understand the nature of these exotic states, i.e., whether they are compact multiquark states or loosely bound states of conventional hadrons or mixtures of both, we are still far away from a unified picture which can explain all their relevant properties. As many of these states are located close to the thresholds of two conventional hadrons, they are conjectured to be molecular states or at least contain large molecular components~\cite{Guo:2017jvc}. Nonetheless, it is not an easy task to distinguish between the molecular picture and the competing interpretations. The reason is obvious. Quantum mechanically, a hadronic state can have all the configurations allowed by its quantum numbers. For instance, unless forbidden by its quantum numbers, a meson can have both $q\bar{q}$ and $qq\bar{q}\bar{q}$ components. The latter can be further  classified into a meson-meson configuration or a compact tetraquark configuration. We refer to a state as a conventional meson if the $q\bar{q}$ component is dominant, or a meson-meson molecule if the meson-meson configuration is dominant.

Recently, it has been proposed that to clarify the nature of the exotic states, such as  $X(3872)$, $D_{s0}^*(2317)$, the $P_c$ pentaquark states, and $T_{cc}^{+}(3875)$ mentioned above, or to either confirm or refute the molecular picture, one can turn to multi-hadron systems~\cite{Wu:2021dwy}.
The argument goes as follows. We know that atomic nuclei, to a large extent, can be treated as bound states of nucleons, i.e., a deuteron is composed of  one proton and one neutron, a triton is composed of one proton and two neutrons, an alpha particle contains two protons and two neutrons,  and so on. Such a picture has been tested extensively. Nowadays, using nucleons as degrees of freedom and with the two-body $NN$ interaction determined by the nucleon-nucleon scattering data and the residual small $NNN$ interaction, ab initio calculations can reproduce most of the ground and low-lying excited states of light- and medium-mass nuclei. Adding hyperons to the nuclear system, one ends up with  hypernuclei. Following the same approach, the properties of hypernuclei can also be understood very well.

Regarding the exotic hadrons that do not fit the naive QM picture, if we believe that one exotic hadron $C$ is a bound state of two conventional  hadrons $A$ and $B$. 
Then with the same interaction between $A$ and $B$ which leads to the formation of $C$, one can study the three-body $ABB$ and $AAB$ systems and check whether they can form bound/resonant states. If the $ABB$ or $AAB$ system binds, then an experimental (or lattice QCD) confirmation on the existence of this state can unambiguously verify the molecular picture for the exotic state $C$, i.e., it is  dominantly a hadronic molecule of $A$ and $B$. 
Conversely, studies on multi-hadron states can also shed light on the relevant hadron-hadron interactions.  It should be noted that the binding of a three-body system is not only caused by the interactions but also other factors such as their masses and quantum number configurations. A low reduced mass can increase the kinematic energy barrier and make the systems more difficult to bind. Quantum number configurations such as spin and isospin operator coupling coefficients in multi-body systems can greatly affect the bindings of multi-hadron systems as well. In the case that $AA$ ($BB$) is not bound, the binding of $AB$ is not a sufficient but necessary condition for the binding of $AAB$ ($ABB$).

In the light $u$, $d$, and $s$ flavor sector, the $\bar{K}NN$ and $\bar{K}NNN$ quasibound states are the most studied multi-hadron states~\cite{Hyodo:2022xhp}. The $\bar{K}N$ interaction  is attractive such that  the $\Lambda(1405)$ state can be understood as a $\bar{K}N$ molecule. This has motivated searches for $K^-$-nucleus quasibound states such as the $\bar{K}NN$ and $\bar{K}NNN$ states. The lightest $\bar{K}$-nuclear quasibound state  $K^-pp$, with isospin $I=\frac{1}{2}$ and spin-parity $J^{P}=0^-$, is dominated by the $I_{\bar{K}N}=0$ $S$-wave interaction. The $\bar{K}NN$ system has been thoroughly studied in various methods, such as the Fadeev equation, variational method and fixed center approximation. 
All these studies find that the $\bar{K}NN$ system can bind with a large width (about 30-90 MeV). The studies employing energy-dependent $\bar{K}N$ interactions in unitary chiral models  yield smaller binding energies of 9$\sim$23 MeV  than the studies employing energy-independent phenomenological interactions (about 40-80 MeV). This may be partly attributed to the higher $\Lambda(1405)$ mass obtained in unitary chiral models. It should be mentioned that several experiments already reported  evidence for the $\bar{K}NN$ quasi-bound state~\cite{Hyodo:2022xhp}.

Another impressive work in light-flavor sector is the multi-$\rho$ study~\cite{Roca:2010tf}, the authors use the $\rho\rho$ interaction in spin 2 and isospin 0 channel to show that the resonances $f_2(1270)$ ($J^{PC} =2^{++}$), $\rho_3(1690)$ ($3^{--}$), $f_4(2050)$ ($4^{++}$), $\rho_5(2350)$ ($5^{--}$) and $f_6(2510)$ ($6^{++}$) are basically molecules of increasing number of $\rho(770)$ mesons. The masses of the multi-$\rho$ states are found very close to the experimental values and an increasing value of the binding energy per $\rho$ as the number of $\rho$ mesons is increased. Another interesting observation is that the width of the multi-$\rho$ state increases with the number of $\rho$ mesons. To the point that for $n_{\rho} = 6$ the width is already very large and reaches the experiment threshold with present detection techniques, which may be responsible for the end of multi-$\rho$ observation.

\begin{figure*}[h!t]
    \centering
    \includegraphics[width=0.8\linewidth]{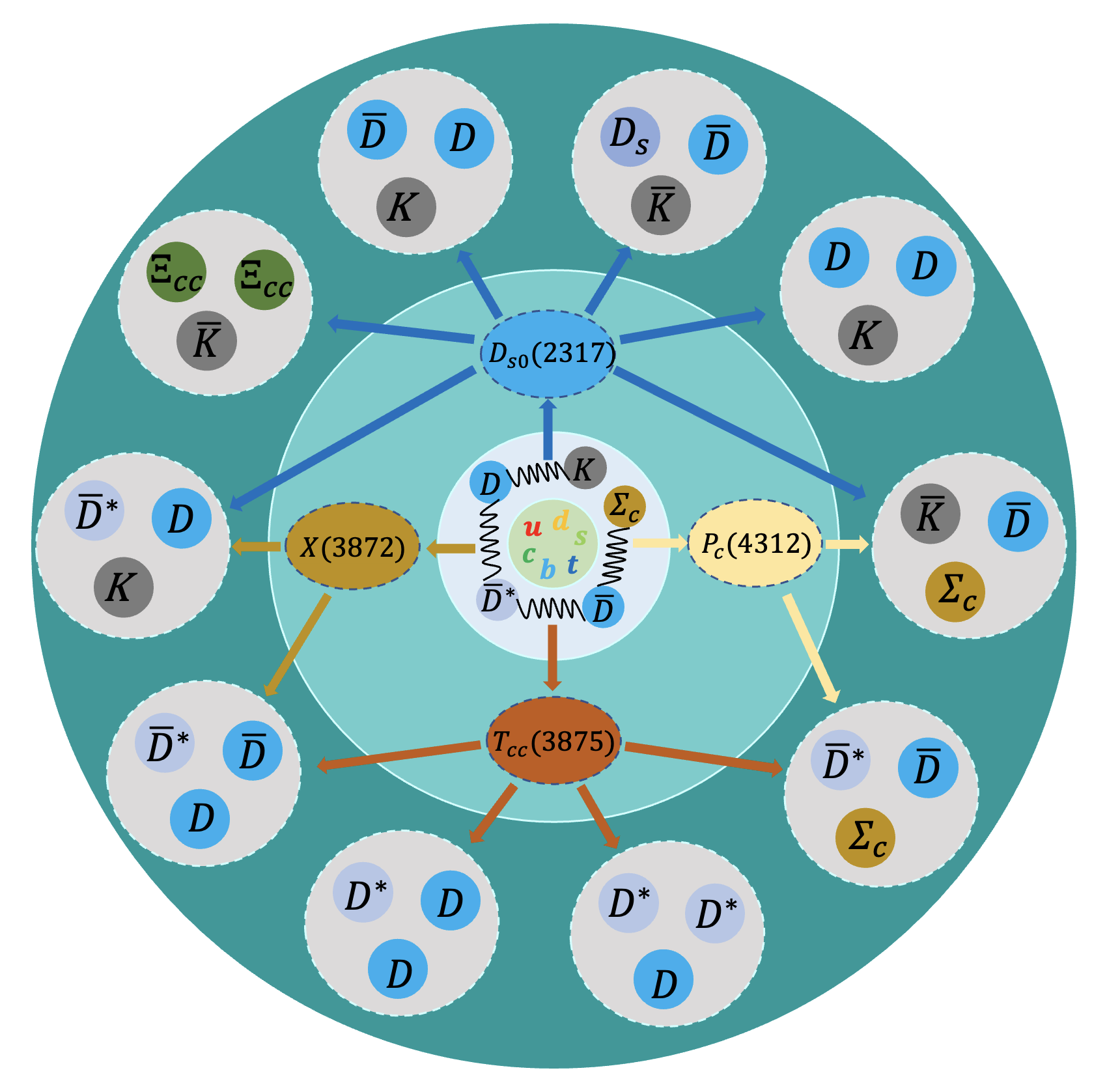}
    \caption{Multi-hadron states built upon the molecular picture where $D_{s0}(2317)$ is a bound state of $DK$, $X(3872)$ is a bound state of $D\bar{D}^*$, $P_c(4312)$ a bound state of $\bar{D}\Sigma_c$, and $T_{cc}(3875)$ a bound state of $DD^*$. From inside to outside, the innermost circle encompasses the six quarks, the next circle encompasses the $q\bar{q}$ and $qqq$ baryons, the following circle encompasses the two-hadron molecules while the outermost circle encompasses the three-hadron molecules.}
    \label{fig:1}
\end{figure*}

More multi-hadron states are expected in the heavy-flavor sector containing charm or bottom quarks  due to the fact that many two-body molecular candidates have already been discovered experimentally and identified theoretically and that more symmetries play a role here, such as heavy quark spin and flavor symmetries, and heavy antiquark diquark symmetry. In Fig.~\ref{fig:1}, we present some of the predicted multi-hadron states based on the molecular picture where  $D_{s0}^*(2317)$, $X(3872)$, $P_c(4312)$, and $T_{cc}^{+}(3875)$ are two-body molecules. For other systems studied from this perspective, see Refs.~\cite{Wu:2021dwy,MartinezTorres:2020hus}.

\begin{widetext}  
 \begin{table*}[h!]
    \centering
    \caption{Binding energies and specific partial decay widths of multi-hadron states in the heavy flavor sector with their sub two-body bound states  and the corresponding hadronic molecular candidates. The Binding energies and widths are in units of MeV.}
    \label{tab:1}
    \begin{tabular}{c c c c c c}
    \hline
    \hline
      Multi-hadron states&$I(J^P)$ & B.E. (MeV) & Decay channels (width) & Sub two-body bound states (B.E.) & Two-body molecular candidates \\
      \hline
       $DDK$& $\frac{1}{2}(0^-)$ & $\sim70$~\cite{Wu:2019vsy}  & \thead{${DD_s^*}$ ($\sim0.3$)\\$ D^*D_s$ ($\sim2.4$)}~\cite{Huang:2019qmw}&$DK$ ($\sim$45)& $D_{s0}^*(2317)$\\

    $D\bar{D}K$& $\frac{1}{2}(0^-)$  & $\sim49$~\cite{Wu:2020job}  & \thead{${D_s\bar{D}^*}$ $(\sim0.5)$\\ $J/\psi$K ($\sim$0.2)}~\cite{Wu:2020job}& $DK$ ($\sim$45)& $D_{s0}^*(2317)$\\
    
       $D\bar{D}^*K$& $\frac{1}{2}(0^-)$  & $\sim77$~\cite{Wu:2020job}  & \thead{${D_s^{(*)}\bar{D}^{(*)}}$ $(\sim1.9)$\\$J/\psi K^*(892)$ ($\sim$6.7)}~\cite{Ren:2019umd}& \thead{$DK$ ($\sim$45)\\$D\bar{D}^*$ ($\sim$4)}& \thead{$D_{s0}^*(2317)$\\$X(3872)$}\\
        
     $\Sigma_{c}\bar{D}\bar{K}$& $1(\frac{1}{2}^+)$& $\sim78$~\cite{Wu:2021gyn}  & \thead{${D\Xi'}$ (
     $\sim90$)\\${D_s\Sigma_c}$ $(\sim20)$}~\cite{Wu:2021gyn}& \thead{$DK$ ($\sim$45 )\\ $\Sigma_{c}\bar{D}$ ($\sim$8.9)} & \thead{$D_{s0}^*(2317)$ \\$P_c(4312)$} \\
    
     $DDD^*$&$\frac{1}{2}(1^-)$ &  $\sim1$~\cite{Wu:2021kbu} &\thead{${DDD\pi}$\\${DDD\gamma}$}~\cite{Wu:2021kbu} & $DD^*$ ($\sim0.3$)& $T_{cc}^{+}(3875)$\\
     
          \hline
          \hline
    \end{tabular}
\end{table*}   
\end{widetext}

\begin{figure*}[h!t]
\begin{center}
\includegraphics[width=7.0in]{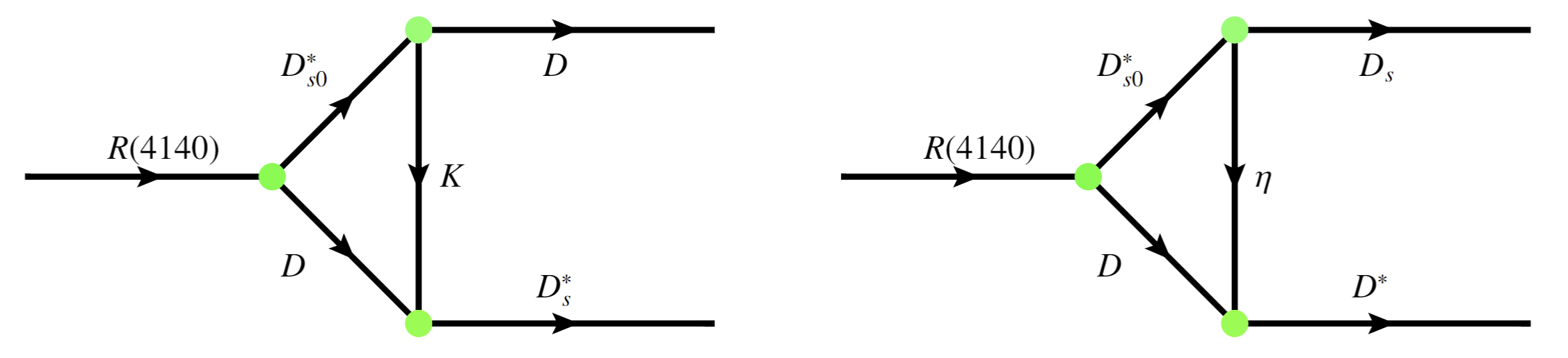}
\caption{Strong decays of the $DDK$ bound state $R(4140)$ to $DD_s^{\ast}$ and $D_{s}D^{\ast}$ via triangle diagrams.       }
\label{decay}
\end{center}
\end{figure*}

In the following, we choose the $DDK$ system that is built upon  $D_{s0}^*(2317)$ as an example to briefly explain these studies. The $D_{s0}^*(2317)$ state was first discovered by the BABAR Collaboration and then confirmed by the CLEO and Belle Collaborations. It is located  45 MeV below the $DK$ threshold and has a decay width less than 3.8 MeV. The observed mass and width are far away from the predicted mass of 2460 MeV and width of hundreds of MeV in the naive quark model. Thus  $D_{s0}^*(2317)$ cannot be a conventional $P$-wave $c\bar{s}$ state. On the other hand, due to the strongly attractive $DK$ interaction, it can  be easily explained as a $DK$ molecule.

It is interesting to note that in the unitarized chiral approach the $D^*K$ interaction is the same as the $DK$ interaction up to heavy quark spin symmetry breaking effects. As a result, the existence of a $DK$ molecule implies the existence of a $D^*K$ molecule. 
In Ref.~\cite{Altenbuchinger:2013vwa}, fixing the next-to-leading order low-energy constants (LECs) and a subtraction constant by fitting to the lattice QCD  scattering lengths, and then solving the Bethe-Salpeter equation, one found two poles in the strangeness 1 and isospin 0 channel, which coincide with the experimentally known $D_{s0}^*(2317)$ and $D_{s1}(2460)$. In such a picture, one can easily understand the fact that the mass difference between $D_{s1}(2460)$ and $D_{s0}^*(2317)$ is almost the same as the mass difference between $D^*$ and $D$ because in the molecular picture, the mass difference originates from the different masses of the constituents $D$ and $D^*$, as the interactions between $D^*K$ and $DK$ are the same due to heavy quark spin symmetry.  In addition, chiral symmetry also dictates that the $\bar{D}(\bar{D}^*)K$ interaction is half that of the $DK(D^*K)$ interaction.

 The strongly attractive $DK(D^*K)$ interaction, inferred from the existence of $D_{s0}^*(2317)$ and $D_{s1}(2460)$ and confirmed by lattice QCD simulations and chiral symmetry, indicates that  the $DDK$, $D\bar{D}K$, and $D\bar{D}^*K$ three-body systems can bind. This has been confirmed by sophisticated few-body studies (e.g., the gaussian expansion method)~\cite{Wu:2019vsy,Wu:2020job}. The influences of the $DD$ and $D\bar{D}$  interactions on the corresponding three-body systems have been studied as well.  The results show that the $DDK(D\bar{D}K)$ system is mainly formed by the $DK$ interaction and the $DD(D\bar{D})$ interaction only affects the binding energies by a few MeV~\cite{Wu:2019vsy,Wu:2020job}. For the $D\bar{D}^*K$ system,  the $D\bar{D}^*$ interaction is strong enough to form the $X(3872)$, and therefore it also plays an important role in generating the $D\bar{D}^*K$ bound state~\cite{Wu:2020job}. An experimental discovery of any of these states can provide a highly nontrivial check on the molecular picture of $D_{s0}^*(2317)$ and of the many other exotic states of similar kind.  It is interesting to note that the predicted $DDK$ multi-hadron state has been searched for by the Belle Collaboration in $e^+e^-$ collisions~\cite{Belle:2020xca} and inspired theoretical studies in finite volume~\cite{Pang:2020pkl}.

 In the same way, one can deduce two-body interactions between the constituents of the $X(3872)$,
 pentaquark states, and $T_{cc}$. With these interactions, one can solve the three-body Schr\"odinger equations (using different methods), and search for bound states. 
 In Table~\ref{tab:1},  some of the predicted multi-hadron molecules  are presented, including their isospin-spin-parity, binding energies, and specific partial decay widths. It should be noted that the existence of these multi-hadron states is tied to the existence of specific  two-body hadronic molecules. Note that in most (if not all) of multi-hadron studies, genuine three-body interactions are neglected. It is well known that the three-nucleon interaction plays an important role in more quantitative understanding of many nuclear phenomena. Nonetheless, for multi-hadron states reviewed here, their existence will not affect qualitatively the results. Once more data become available in the future, the impact of three-body interactions should be studied.  Clearly,  these states are just the tip of the iceberg of possible multi-hadron states that may exist, because  there are  more degrees of freedom in the heavy flavor sector than in the nucleon-nucleon/hyperon sector.

The next important step forward in studies of multi-hadron states is to study their productions and decays. That is to say, where and how to search for them. In Ref.~\cite{Huang:2019qmw}, the partial decay widths of the $DDK$ three-body bound states were studied in the triangle mechanism, see Fig.~\ref{decay}. In Ref.~\cite{Ren:2019rts}, the possibility to observe the $D\bar{D}^*K$ state in $B$ decays was investigated.  The decays of some other multi-hadron states have also been studied in the triangle formalism, such as the $D\bar{D}K$~\cite{Wu:2020job}, $D\bar{D}^{*}K$~\cite{Ren:2019umd}, and $\Sigma_c\bar{D}\bar{K}$~\cite{Wu:2021gyn} states. The results are summarized in Table~1. It should be noted that there is an important difference between multi-meson  and multi-baryon states. For multi-baryons (e.g. the well-known nuclei), due to the baryon number conservation, the decay phase space is suppressed, while for multi-mesons, no such conservation law exists, which means that a $N$-meson state could decay into $N'$($<N$) mesons and may cause the decay width of multi-meson states too large to be observed experimentally. Even though, there exist some universal properties. The nature of baryon number conservation is that the strong interaction does not change the quark flavor, which is  true for both mesons and baryons. Thus we can use the minimum quark content  to label the multi-quark/multi-hadron states.  The nature of a $N$ meson state decaying to $N'$ mesons is  quark-antiquark annihilation/production.
For example, for the strong decay of the $DDD^*$ bound state ($ccc\bar{q}\bar{q}\bar{q}$),  the final particles should be no less than three mesons (or two baryons). And if the multi-hadron states are bound states of ground-state particles,  the phase space is also suppressed.
However, in general, the decay phase space of multi-meson bound states indeed is larger than that of multi-baryon bound states. Clearly more studies are needed to provide guidance for future experimental searches.

It is interesting to note that $pp$ collisions, heavy ion collisions, and even $e^+e^-$ collisions  are able to produce a large amount of conventional hadrons. When they are close enough, the residual strong interaction may  bind them such that multi-hadron molecules can be produced. There have been a number of studies along this direction on the two-body hadronic molecules~\cite{Zhang:2020dwn}. It will be extremely useful that  studies on heavy-flavor multi-hadron molecules are performed along the same line. 

P. W. Anderson once said, ``more is different", which could also be true in hadron physics. Studies of  multi-hadron molecules have just started and are in an infant stage, compared with the studies of multi-nucleon states (nuclei) and  of two-body hadronic molecules.
There is no doubt that the studies of multi-hadron states  will significantly improve our understanding on their nature as well as the non-perturbative strong interaction in general. 
As a result, we envisage a bright future for such studies both experimentally and theoretically and expect to see more theoretical, experimental, and lattice QCD studies on multi-hadron molecules at present high energy facilities such as LHCb, BEPC, and KEKb and planned facilities such as high luminosity LHC or super Tau-Charm  in the future.

%


\section*{Conflict of interest}
The authors declare that they have no conflict of interest.

\section*{Acknowledgments}
This work is supported in part by the National Natural Science Foundation of China under Grants No.12147152, No.11735003, No.11975041, and No.11961141004.

\bibliography{multihadron}

\end{document}